\documentclass[twoside,10pt]{article}
\usepackage{graphicx}

\title{\Large \bf Charge exchange reaction at high energies}
\author{R. Schicker$^1$, L. Jenkovszky$^2$, O. Kuprash$^3$}
\date{}
\begin{document}
\maketitle

\begin{center}
\vspace*{-0.3cm}
{\it  $^1$~Phys. Inst., Heidelberg\\
$^2$~BITP, Kiev\\
$^3$~DESY, Hamburg}
\end{center}

\vspace{0.3cm}

\begin{center}
{\bf Abstract}\\
\medskip
\parbox[t]{10cm}{\footnotesize
The charge exchange reaction pp $\rightarrow$ n $\Delta^{++}$ at high energies 
is examined. The cross section of this reaction is estimated for RHIC and LHC
energies based on data taken at the lower energies of the Intersecting 
Storage Ring (ISR) at CERN. The interest of such measurements for identifying
the associated electromagnetic bremsstrahlung radiation is discussed.}
\end{center}

\section{Introduction} \label{s1}

Diffractive reaction channels contribute about 30\% to the total hadronic
cross section at the energies of the Large Hadron Collider LHC at CERN.
A good understanding of hadronic diffraction is therefore necessary for a 
comprehensive understanding of proton-proton collisions. In single and double 
diffractive dissociation, one or both of the protons get diffractively excited,
and the proton remnants are very much forward focussed\cite{Jenk}. 
In central diffraction, a hadronic system is formed at mid-rapidity
by the fusion of two Pomerons. The above reactions can, however,
also be initiated by the exchange of a photon or a Reggeon.
Such non-diffractive exchanges represent a potential background in the 
analysis of diffraction at high energies. 
A good understanding of electromagnetic processes, as well as of Reggeon 
exchanges, is therefore mandatory. The analysis of reaction channels which 
are purely photon or Reggeon induced are therefore of interest for identifying
possible background sources in diffraction.
 
\section{Charge exchange reactions} \label{sec2}

The study of charge exchange mechanism is of interest for an improved understanding
of purely Reggeon induced reactions. Reggeon trajectories are defined 
by bound states of $q\overline{q}$-pairs. The QCD content of these exchanges was 
studied in Ref.\cite{Lipatov1}, and remains a challenge for theory.
The exchange of a charged $q\overline{q}$-pair represents
a sudden acceleration of electric charge, accompanied by electromagnetic 
bremsstrahlung radiation. The measurement of this radiation is of interest
as a test of the theorem of Low for soft photon emission as discussed below.

A variety of final states is available in charge exchanges in proton-proton 
collisions. Charge exchange reactions can be due to the exchange of the pion-,
$\rho$- or A$_{2}$-trajectory. 

\begin{eqnarray}
pp \rightarrow n + \; \Delta^{++} \rightarrow n + p\pi^{+} \\
pp \rightarrow \Delta^{0} \;+ \; \Delta^{++} \rightarrow n\pi^{0} + 
p\pi^{+} \\
pp \rightarrow \Delta^{0}  \;+ \; \Delta^{++} \rightarrow p\pi^{-} + p\pi^{+}
\label{eq:01}
\end{eqnarray}

The simplest approach for studying these charge exchange reactions
is in binary kinematics.

\begin{figure}[h]
\begin{center}
\includegraphics[width=1.0\textwidth]{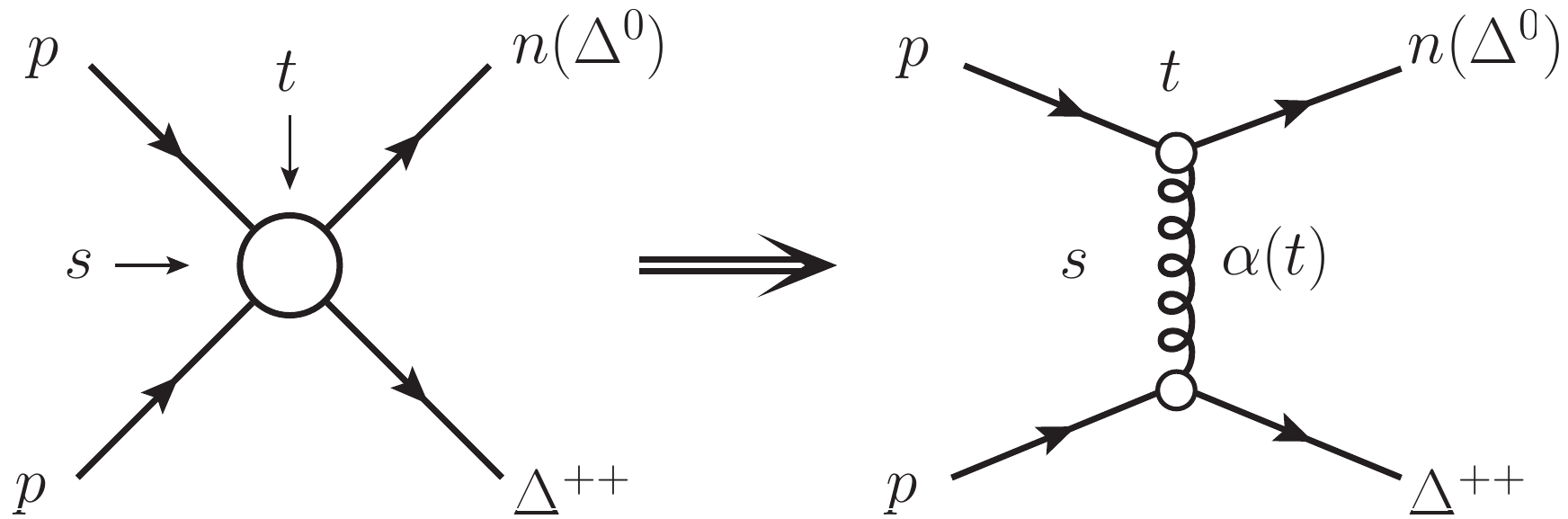}
\end{center}
\vspace{-0.5cm} \caption{Charge exchange reaction in binary kinematics.} 
\label{fig:1}
\end{figure}

The relevant scattering amplitude at high energies, with a single Regge 
exchange (pion, $\rho$, A$_{2}$) is

 \begin{equation}\label{ampl}
    A^{\pm}(s,t)=A_i\xi^{\pm}(t)\beta(t)(s/s_i)^{\alpha(t)},
 \end{equation}
with $A^{\pm}$ the amplitude for l=even and l=odd exchange, respectively.
The variables  $s=(p_1+p_2)^2=(p_3+p_4)^2,\ \ t=(p_1-p_3)^2=(p_4+p_2)^2$ are the usual 
Mandelstam invariants, $\xi^{\pm}(t)$ is the signature factor 
($\xi^{\pm}(t)=1 \pm e^{-i\pi\alpha(t)}$), and $\beta(t)$ is the residue function 
taken in exponential form $\beta(t)=e^{b_it}$, where $b_i$ is fitted to each reaction, 
$pp\rightarrow n\Delta^{++}$  or $pp\rightarrow \Delta^0\Delta^{++}$. In this type of 
reactions, the dominant exchange is the pion trajectory at low energies with the 
$\rho$-trajectory dominating at high energies. We take a linear form 
$\alpha(t)=0.0+0.8t$ and $\alpha(t)=0.5+0.9t$ for the pion and $\rho$-exchange, 
respectively\cite{Kaidalov1}. 

Apart from the overall normalization parameter and the standard Regge scale 
factor $s_i$, that we set equal to $1$ GeV$^2$, the model contains only one 
free parameter, namely $b$ that we find from comparison with measured data. 

With the a priory unknown normalization

\begin{equation}\label{norm}
\frac{d\sigma}{dt}=\frac{\pi}{s^2}|A(s,t)|^2
\end{equation}
one can calculate the cross section. By a comparison with the existing data 
and fits, e.g.\cite{Kaidalov2}, one finds the differential and integrated 
cross section.

Measurements of the reaction shown in eq.1. above were made at the Argonne 
National Zero Gradient Synchrotron\cite{Argonne}, and at the Intersecting 
Storage Ring at CERN\cite{ISR}.

The analysis of the cross section of pp $\rightarrow$ n$\Delta^{++}$ at 
momenta $P_L$ up to \mbox{11 GeV/c} finds good agreement with the Chew-Low 
one-pion-exchange mechanism taking into account form-factor 
models\cite{Chew,Form1,Form2}.

The analysis of the ISR data identifies the $\rho$- and 
A$_{2}$-contribution which start to dominate the cross section at energies 
$\sqrt s$ = 31 GeV\cite{Moriarty}.    

In order to evaluate the cross section at RHIC and LHC energies, we take the 
ISR data at $\sqrt s$ = 31, 45 and 53 GeV and fit the data as outlined above. 


\begin{table}[h] \noindent\caption{Cross section pp $\rightarrow$ n$\Delta^{++}$}
\vskip3mm\tabcolsep5pt

\begin{center}
{\begin{tabular}{|c c c |}
 \hline \multicolumn{1}{|c}{} &\multicolumn{1}{|c}{$\sqrt s$ (GeV)} 
&\multicolumn{1}{|c|}{$\sigma$ (nb)}\\%
\hline%
ISR & 31 & 580 $\pm$90 \\
& 45 & 210$\pm$40 \\
& 53 & 170$\pm$40 \\ \hline
RHIC & 100 & 48.5$\pm$5.5\\
& 200 & 12.2$\pm$1.3 \\ \hline
LHC & 7$\times 10^3$ &  (10.0$\pm$1.1) $\times 10^{-3}$ \\
& 14$\times 10^3$ & (2.4$\pm$0.3) $\times 10^{-3}$\\
\hline
\end{tabular}}
\end{center}
\end{table}

The cross sections extrapolated to the RHIC energies of $\sqrt s$ = 100 and 
\mbox{200 GeV} shown in Table 1 are 48.5 and 12.2 nb, respectively. The 
corresponding values for the LHC energies of $\sqrt s$ = 7 and 14 TeV are 
10.0 and 2.4 pb, respectively. These values can be used for further evaluating 
experimental prospects of measuring the charge exchange channel pp 
$\rightarrow$ n$\Delta^{++}$ at these collider energies.

\section{Electromagnetic radiation in charge exchange reactions} \label{sec3}

Accelerated electric charge emits bremsstrahlung radiation\cite{Jackson}.
The theorem of Low relates the radiative leading and next-to-leading order in 
photon energy of the bremsstrahlung amplitude to the corresponding 
non-radiative amplitude shown in Fig.\ref{fig:1}\cite{Low}. 
The photons emitted from the external lines result in a pole term
in the radiative amplitude, and generate the leading $k^{-1}$-dependence
in the photon energy spectrum. Emission from the internal lines
gives rise to the next-to-leading constant term in the energy spectrum. 
The theorem of Low applies to photons in the soft limit, i.e. to photon energies
which are smaller than any other momentum scale in the amplitude.
The emission of soft photons in the high energy limit was studied
in Ref.\cite{DelDuca}. This study finds correction in the next-to-leading
radiative amplitude due to the internal structure of the external states.
A generalization of the theorem of Low to the production of non-abelian gauge 
bosons and gravitons is presented in Ref.\cite{Lipatov2}. 

\section{Summary and outlook}

The cross section for the charge exchange reaction pp $\rightarrow$ n$\Delta^{++}$ is 
extrapolated from data at the ISR to the energies of RHIC and LHC. More refined 
calculations are under way, and will take into account the three body nature of the final 
state, including the non-resonant continuum at masses beyond the \mbox{$\Delta$-resonance.}
The interest for measuring the associated bremsstrahlung photons is outlined. 
Quantitative predictions for this  associated radiation are under study.

\section{Acknowledgments}

We gratefully acknowledge fruitful discussions with L. Lipatov during
the Crimean Conference on issues related to the subject of this paper.
This work is partially supported by WP8 of the hadron physics program of 
the 8th EU program period.

\end{document}